\documentclass[a4paper,10pt]{article}

\usepackage[utf8x]{inputenc}
\usepackage{natbib}
\usepackage{authblk}
\usepackage{hhline}
\usepackage{graphicx}

\title{Atom probe analysis of ex-situ gas-charged stable hydrides}
\author{D. Haley, P.A.J. Bagot, M.P. Moody}
\affil{Dept.\ of Materials, University of Oxford, Parks Rd, Oxford, Oxfordshire, United Kingdom, OX1 3PH}
\date{}

\begin{document}
\maketitle

\section{Abstract}
In this work we report on the atom probe tomography analysis of two metallic hydrides, formed by pressurised charging using an ex-situ hydrogen charging cell, in the pressure range of 200-500~kPa (2-5~bar). Specifically we report on the deuterium charging of Pd/Rh and V systems. Using this ex-situ system, we demonstrate the successful loading and subsequent atom probe analysis of deuterium within a Pd/Rh alloy, and demonstrate that deuterium is likely present within the oxide-metal interface of a native oxide formed on vanadium. Through these experiments, we demonstrate the feasibility of ex-situ hydrogen analysis for hydrides via atom probe tomography, and thus a practical route to 3D imaging of hydrogen in hydrides at the atomic scale.

\section{Introduction}
The characterisation of the atomic-scale behaviour of hydrogen remains a challenging task for high resolution microscopy techniques. This is increasingly restrictive to the further development of new materials whose optimisation critically depends on a more thorough understanding of how individual hydrogen atoms will distribute in the material’s microstructure during service. In particular, this is the case in hydrogen embrittlement, where under load materials can suddenly fail at elongations well below their normal service levels. 

This sudden failure is a problematic phenomenon for the safety and performance of structural materials, and is pervasive across a wide range of industries. An improved capability in hydrogen imaging has implications not only to limiting the effects of embrittlement, but also in advancing research into fuel-cell energy storage applications~\citep{durbin2013review}. 

Much of the current information regarding the interaction of hydrogen with metallic systems has been obtained using macroscopic characterisation methods. This is in contrast with standard analysis methodologies for higher-Z elements, whereby microscopic methods can provide spatially resolved insight into elemental distributions, and the associated macroscopic behaviour. Typical characterisation methods that utilise photon or electron radiation have only a weak interaction with the hydrogen atom owing to its small charge cloud and thus it is not readily detected. Recent advances in Transmission Electron Microscopy (TEM) have enabled the imaging of hydrogen atoms under very specific conditions~\citep{Meyer2008}, and in some pure crystalline compounds~\citep{Ishikawa2011} at high concentrations. Arguably, the most applicable methods in microscopic hydrogen analysis are Secondary Ion Mass Spectroscopic (SIMS)~\citep{Carlson1978} or Nuclear Reaction Analysis (NRA) and the related Elastic Recoil Detection (ERD) techniques~\citep{Lanford1992}.

These techniques are ion beam-based methods and are restricted to near-surface zones, with penetration depths on the micron-scale. Both ERD and NRA are highly specialised techniques. NRA in particular requires high energy incident radiation from specific isotopes, for example ${}^7\mathrm{Li}$: 3.0~MeV, ${}^{15}\mathrm{N}$: 6.4~MeV. Access to such high-energy isotope sources are generally not available to typically sized materials laboratories.

Atom Probe Tomography (APT) is emerging as a candidate technique for the study of hydrogen. APT is able to analyse the distribution of H by examining ${}^2\mathrm{H}$ (Deuterium) isotopes. This approach is enabled by the natural scarcity of deuterium, 100~ppm ${}^2\mathrm{H}$ per H, for isotopic labelling to unambiguously distinguish real signal from the otherwise omnipresent H signal that is detected in the instruments high vacuum environment. Previous APT studies of hydrogen have explored hPa-level hydrogen atmospheres~\citep{Gemma2012,Takahashi2010,Takahashi2012}, and more recently the introduction of hydrogen electrolytically into materials~\citep{Haley2014}.

To date, there have been few investigations of examinations of hydrides formed by atmospheric or pressurised charging. Such hydrides are of interest in energy storage applications, and the experimental feasibility in APT for these types of material has not yet been demonstrated. This research demonstrates both the feasibility and the limitations of performing deuteration studies of hydrides via ex-situ deuterium charging, an accessible method for existing APT laboratories.

\section{Materials and Methods}
Two hydride forming materials were investigated in this study. Firstly, a Pd-Rh alloy is studied. This class of alloys has applications for catalysis~\citep{Li2012} and notably is useful in converting $\mathrm{H_2}$ to its atomic form, enabling subsequent absorption by other materials~\citep{Konda2015}. Secondly, pure vanadium is examined where, at the pressures used in this study, both H and D are in solid solution.  Vanadium is unusual in that the phase diagram for V-H is markedly different to V-D~\citep{Manchester2000}. Thus care must be taken in interpretation of the D interaction within V and the extrapolation of these results to predict the behaviour of H.  For the purposes of this work, these materials are used to demonstrate the capability of APT to characterise hydrides.

The Pd-H system, for which extensive data is available and for which the Pd-D system exhibits similar behaviour, has a unique two-phase region. At room temperature the two-phase $\alpha$-$\beta$ region occurs between 2 and 40 at.\% hydrogen, between which the pressure-composition isotherm is relatively flat~\citep{Manchester2000}. The Pd-Rh-H system behaves similarly, and Rh is considered to not alter D solubility considerably within the pressure range in this study~\citep{Thiebaut1995}. This two-phase region is slightly wider at the cryogenic temperatures used by APT analysis, but otherwise should not differ for the purposes of this work.

Solubility isotherm data indicates that at equilibrium with 100~kPa (1~bar) hydrogen, the equilibrium solubility should exceed 0.75~H atoms per metal atom for both Pd and Pd-Rh.  The bulk diffusion rate for Pd-D at room temperature is approximately $5.5 \times 10^{{}-11}$~$\mathrm{m}^2/\mathrm{s}$~\citep{flanagan1991palladium}, which yields a 95\% diffusion time of $5\times10^{{}-5}~\mathrm{s}$ for a 100~nm linear diffusion path, indicating rapid equilibriation within the bulk. However, adsorption/desorption kinetics of H to/from the surface of the material are likely to dominate the time to equilibriation~\citep{Schwarz2005}. Thus these time-scales are likely to be widely underestimated in describing actual diffusion behaviour.

Pure V (Goodfellow, 99.8\%) and an alloy of Pd-6.25\%Rh (Johnson Matthey) both in the form of drawn wires were used in this study. The needle-shaped specimens required for APT were generated using standard electropolishing techniques~\citep{Miller1996}. For V a sulfuric-ethanol solution was used, whereas, a two-stage 25\% perchloric-75\% acetic, then 10\% perchloric-acetic mixture was used for Pd-Rh. 

Prior to deuterium charging, samples were initially subject to APT analysis using laser-pulsing mode to remove surface oxides using an incident laser energy of 0.2~nJ, as laser mode has improved yield. APT analysis was then switched to voltage-pulsing mode to obtain a baseline signal in the undeuterated specimens, incorporating approximately $5\times10^5$ detected ions. Samples were then removed from the vacuum, charged with deuterium and re-introduced into the atom probe vacuum system, with air-transfer times on the order of minutes. Post-charging, sample re-introduction into the cryogenically cooled atom probe analysis chamber was achieved within 45~mins from deuterium charge completion.

All APT characterisations were conducted using a reflectron-equipped LEAP 3000HR system in voltage-pulsing mode at 50~K with a 15~\% pulse fraction. Pd data was obtained at a pulse frequency of 100~kHz to prevent Pd mass `wrap-around' (i.e. ions that are incorrectly assigned a subsequent pulse, and thus incorrect time-of-flight due to their long flight time) at low voltages, and V data was obtained at 200~kHz. A voltage reconstruction algorithm (Cameca IVAS 3.6.6) was used to generate 3D data from the APT experiment. 

\begin{figure}[htp]
 \centering
 \includegraphics[width=0.95\textwidth]{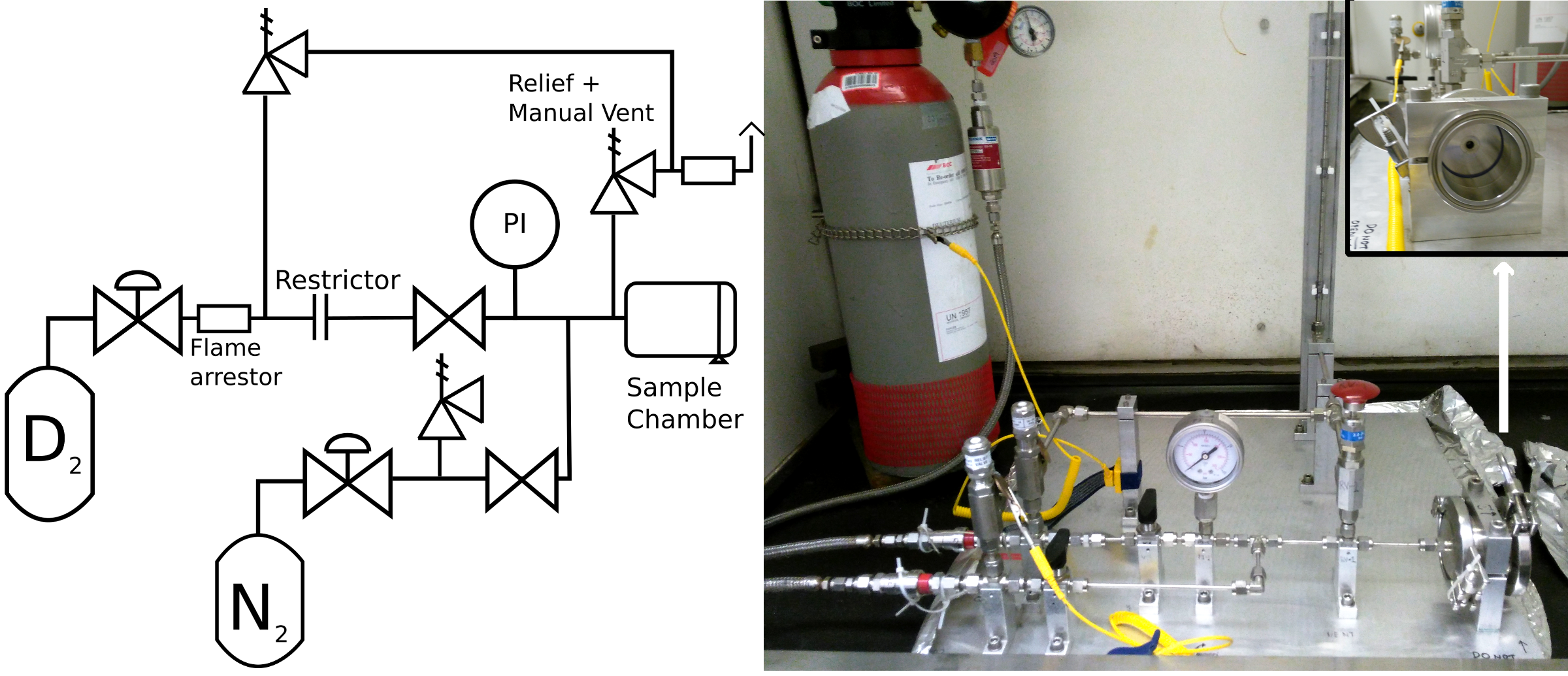}
 \caption{Instrumentation diagram and photograph of the charging rig developed for this study. The charging chamber in the image is at the bottom right. APT samples in holder are inserted and the chamber is pressurised. Deuterium cylinder is shown in the left side of the photograph.}
 \label{fig:ChargingRig}
\end{figure}

A schematic and photograph of the constructed deuterium gas charging system are shown in Figure~\ref{fig:ChargingRig}. The apparatus is a pneumatic system primarily constructed of 316 stainless steel, which allows for $\mathrm{N}_2$ purging and then subsequent deuterium charging of APT samples. The maximum pressure of the system is limited to 500~kPa (5~bar).

APT samples were loaded into the deuterium charging rig whilst still within their APT specimen holders for straightforward transfer to/from the LEAP 3000HR. Atmospheric oxygen was purged by at least 5 cycles of purging gas, $\mathrm{N}_2$, at 500~kPa (5~bar) then the pressure reduced to 100~kPa (1~bar) (abs). Next, $\mathrm{D}_2$ was pressurised at 500~kPa (5~bar), then vented to 120~kPa (1.2~bar). This was repeated three times to purify the $\mathrm{D}_2$ content in the chamber. Subsequently gaseous charging with $\mathrm{D}_2$ gas was conducted. After charging, $\mathrm{D}_2$ was vented from the system and then a further purging cycle was conducted to safely remove deuterium gas. Charging times were 1~hr for Pd-Rh, in order to ensure sufficient time for diffusion of D into the sample. For V, the exposure time was 30~minutes, as the diffusion coefficient of D within V at room temperature is somewhat higher,  approximately $1 \times {10}^{-10}~\mathrm{m}^2/\mathrm{s}$. 

Three charging conditions for Pd-Rh were examined, 200~kPa, 500~kPa, and 500~kPa with an additional in-vacuum holding delay of 6~days, to examine the stability of the formed Pd-D over extended periods, and at different pressures. For V, charging was performed at a single pressure of 500~kPa.

\section{Results}
\subsection{Palladium Alloy}
Table~\ref{tbl:PdComposition} presents the uncharged composition for the three experiments for the Pd-Rh alloy. It shows that the primary fluctuation in the measured composition is in the oxygen and Pd content. This is likely due to specimen preparation.

\begin{figure}[htp]
 \centering
  \includegraphics[width=0.95\textwidth]{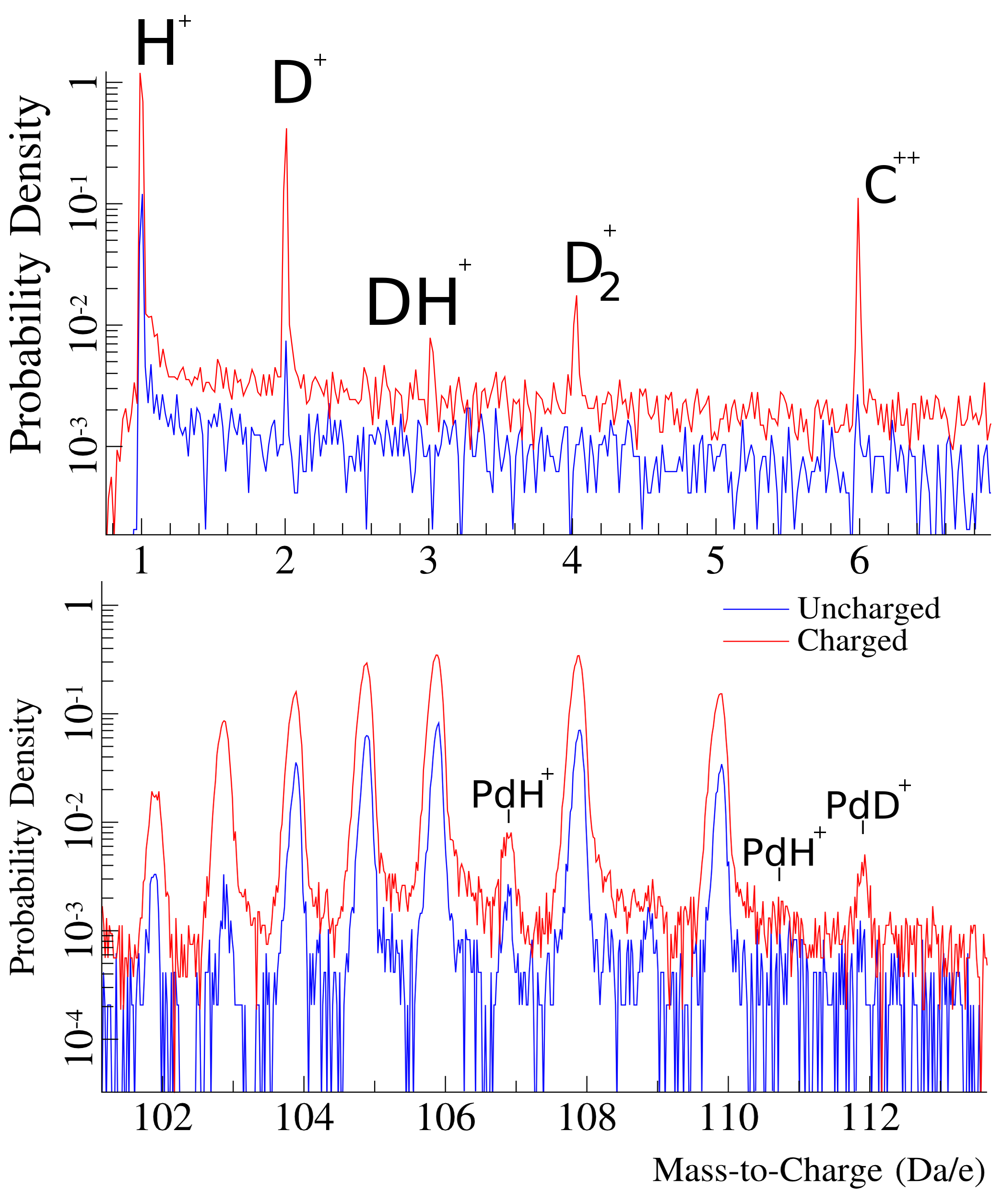}

 \caption{Comparison of charged (500~kPa abs.) and uncharged Pd-6.25\%Rh alloy (same needle), showing peaks at 2 and 4 in the charged state. Normalisation is by m/c spectrum total area.}
 \label{fig:PdChargeComparison}
\end{figure}

After the charging process some surface contamination was found to be present, presumably from the industrial-grade nitrogen source. However this contamination is restricted to the approximately first $1 \times 10^5$ ions of surface material. Comparative pre- and post-charging mass spectra, focussing on the low-mass region and around the $\mathrm{Pd}^{{}+{}}$ isotope peaks are presented in Figure~\ref{fig:PdChargeComparison}.  Note that the comparison spectra were obtained from the exact same atom probe needle.

Figure~\ref{fig:PdChargeComparison} shows that peaks due to D containing species are visible in the post-charging mass spectrum at $m/n$ 2, 3 and 4. In Table~\ref{tbl:PdDCounts}, the measured H:\{$\mathrm{H}_2$,$\mathrm{H}_3$,$\mathrm{H}_4$\} ratios obtained in the uncharged dataset are presented. These ratios will depend strongly on the intensity of the electric field applied during the APT analysis. Noting that the sample should not appreciably change shape due to H exposure since Pd-Rh is a relatively noble alloy, then the field at which the experiment is conducted should be similar pre- and post-charging with D. However, in the case of the charged samples some of these peaks will also contain contributions from D. Thus using the H:\{$\mathrm{H}_2$,$\mathrm{H}_3$,$\mathrm{H}_4$\} ratios  from the uncharged samples ,the $\mathrm{H}_2$ component of the D and the $\mathrm{H}_3$ component of the DH peaks, at 2 and 3~Da respectively, in the analysis of the charged specimen can be estimated and is presented in Table 2.  The primary contaminants detected in the analyses were: C, some unidentified peaks in the range of 13-46~Da, and trace amounts of Hg. 

Deuterium is detected not only in the form of D and $\mathrm{D}_2$ at low mass peaks, but also as a molecular ion $\mathrm{PdD}^{{}+{}}$ in the APT experiment. $\mathrm{PdH}_2$ molecular ions, which would be indistinguishable from PdD, are not considered to have formed as these were not detected within the analysis of the uncharged dataset. Examining Figure~\ref{fig:PdChargeComparison}, PdD formation can be readily seen via ${}^{110}\mathrm{Pd}^2\mathrm{H}$, which has no overlap and thus can be uniquely identified. As a slightly more complex peak, ${}^{105}\mathrm{Pd}^2\mathrm{H}$ only overlaps with ${}^{106}\mathrm{Pd}^1\mathrm{H}$, and is thus separable from Pd at mass 107. ${}^{108}\mathrm{Pd}{}^{1}\mathrm{H}$ and ${}^{110}\mathrm{Pd}{}^{1}\mathrm{H}$ can be unambiguously identified at mass 109 and 111.  As there exists, in general, overlap between Pd, PdH and PdD peaks, numerical separation of the relative contributions~\citep{Miller1996} of these components to each peak must be performed based on natural isotopic abundances of the individual elements that constitute the ion. 

Table~\ref{tbl:PdDFraction} shows the relative quantity of D within each peak from the overlapped Pd/PdH/PdD signal for three sets of charging experiments and subsequent analyses. The applied peak decomposition uses a flat-TOF based algorithm on a nearby peak-free region (90-96~Da) to estimate background noise, and a least-squares non-negative solver for fitting. Statistical testing for non-normality (Anderson-Darling test~\citep{NationalInstituteofStandardsandTechnology2001}) was performed to ensure normality of the background histogram, as used in a previous work~\citep{Haley2015}.

Concentration data for each analysis is reported in Table~\ref{tbl:ChargedPdComposition}. The most significant contaminant species are oxygen, presumably from sample transport, and Hg, which is assumed to be a volatile contaminant in the industrial-grade purging gas. Some unidentified contaminants, possibly organic components from pneumatic lubricants within the charging device were detected, but are not examined for the purposes of this work.
As indicated in Table~\ref{tbl:ChargedPdComposition}, the decomposed deuterium content was variable, between 0.9 and 3~at\%, with slightly higher D contents in the 500~kPa (5~bar) charging. However, this is radically lower than the thermodynamically predicted amounts of D in the bulk phase (40-45~at.\%) for pure Pd.  There is some change in D after holding for 6 days, however as a single data point this result is not statistically robust. Nevertheless, it does demonstrate the stability of the formed hydride within the time and length scales of the APT experiment.

\begin{figure}[htp]
 \centering
   \includegraphics[width=0.95\textwidth]{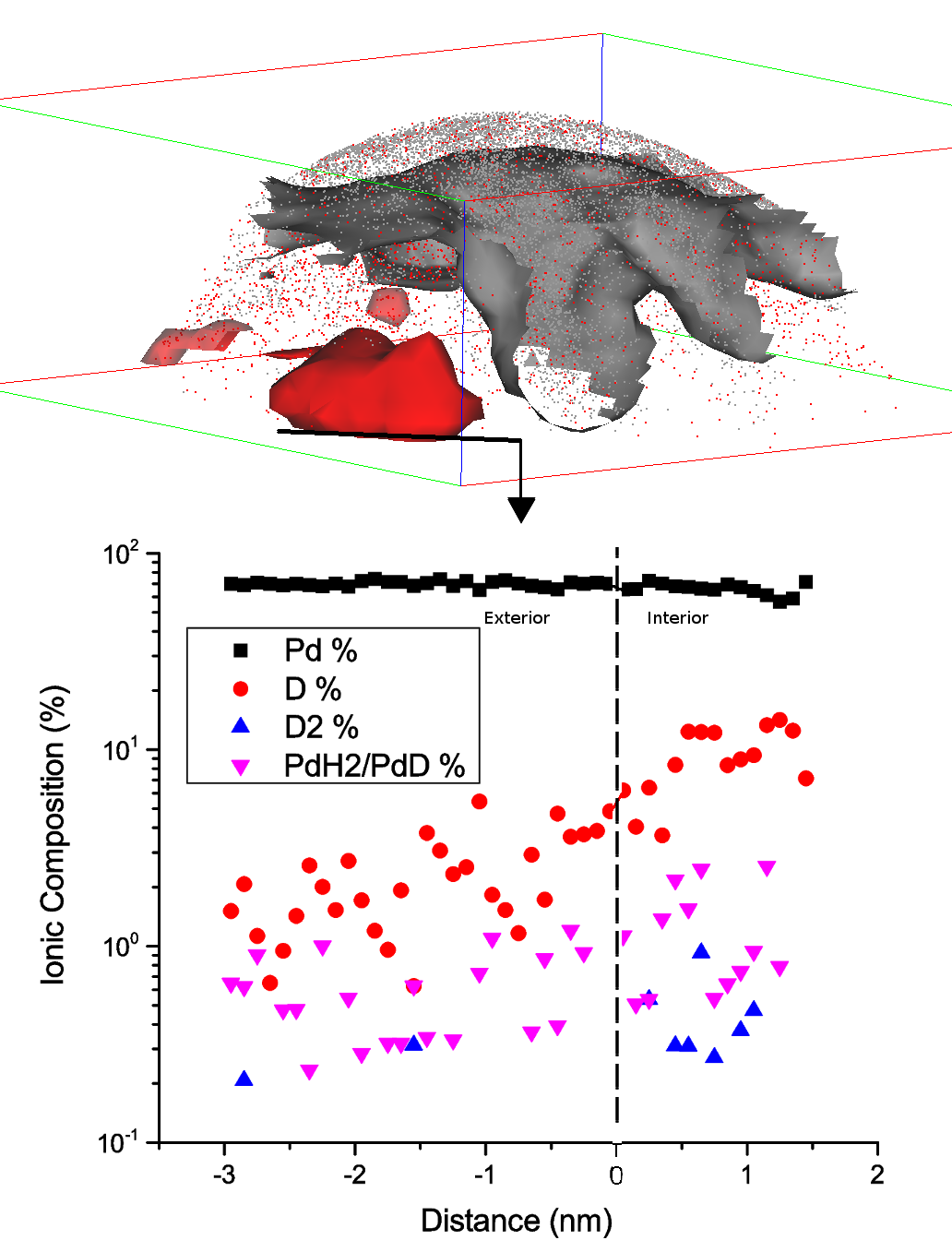}

 \caption{5\% (ionic) D and H isoconcentration surfaces, showing enhanced D zone within the dataset and proxigram of that zone. D concentration increases to up to 10\% D within the marked zone, H is present primarily as a surface artefact. Dataset is not calibrated, so dimensional data should be considered approximate. Note composition scale is logarithmic.}
 \label{fig:PdDProxigram}
\end{figure}

As the system under investigation is within a two-phase region, one could reasonably expect to observe some spatial inhomogeneity within the distribution of D within the atom probe reconstruction due to remaining $\beta$ phase. Indeed a small localised region increased in D is identifiable, within which there is no corresponding increase in H. Figure~\ref{fig:PdDProxigram} shows a reconstruction of a post-charged sample, within which two isoconcentration surface analyses have been applied. The first (red) identifies regions within which the concentration of D is at least 5 at.\%, whereas the second (grey) highlights regions in the reconstruction which the concentration of H is at least 5at.\%. Note, that there is no apparent relationship between the two regions. As shown in the proximity histogram (proxigram), which measures the chemistry as a function of distance from the interface defined by the isoconcentration surface, there is a clear relative increase in D concentration up to 10 at\% D content in the core of the highlighted region. This may be a remnant of some $\beta$ within the material, as it is clear that there is no preference for increased evaporation of H, which would be an artefact of the APT analysis, and only D which is real signal from the charging process.

\subsection{Vanadium}

The initial acquisition of vanadium APT data was performed using laser-pulsing mode to remove surface oxide from the specimen. After removal of the oxide cap, the experiment was then switched to voltage-pulsing mode. The results for the voltage-pulsing are shown in Table~\ref{tbl:ChargedVComposition} for two uncharged runs. Hydrogen and oxygen are the two major contaminant species. Hydrogen was detected as H and $\mathrm{H}_2$ within the sample.

\section{Discussion}

\subsection{Palladium-Rhodium alloy}

From Table~\ref{tbl:PdDCounts}, a snapshot of the D:H ratios that occur during APT field evaporation can be examined. We can further separate the relative contributions within these total signals, i.e. D and $\mathrm{D}_2$ and H and $\mathrm{H}_2$. The estimated ratios of these relative contributions are ~17.7 for H:$\mathrm{H}_2$ and 57.2 for D:$\mathrm{D}_2$. The values of the two measurements are within the same order of magnitude but the significant difference is surprising if the ionisation process is assumed to occur at a fixed conversion rate.  The presence of H is clearly an artefact with respect to the charging process, and can only originate from either within the high vacuum chamber or as an adsorbed species during specimen transfer from the rig to the atom probe. The difference in the measured H:$\mathrm{H}_2$ and D:$\mathrm{D}_2$ ratios may simply be a reflection of this.  

The Pd data in mass spectra obtained from the post-charge state is well explained by fitting the peaks to a mixture of contributions from Pd, PdH and PdD ions.  The results in Table~\ref{tbl:PdDFraction} show that PdH ions do not seem to occur in the APT experiment in large quantities, and that primarily PdD is observed. Significant RhH and RhD contributions seem unlikely, owing to the very good fit to the observed peak amplitudes that can be obtained using only the Pd based isotopes.  The inability to detect PdH ions cannot be directly attributed to a lack of Pd-H surface interaction, but is more likely simply due to the excess D that has been introduced as compared to the background H level. From the presence of D, $\mathrm{D}_2$ and PdD, it is clear that during the charging process D has been taken up successfully in each case.

The relative concentrations of D, as shown in Table~\ref{tbl:ChargedPdComposition}, are significantly below the thermodynamically expected values, even if the H component, an artefact of the experiment, is included in the total concentration 

Several possibilities to explain this observation exist: 

\begin{enumerate}

\item The D concentration has been underestimated due to unidentified species with significantly changed D content due to highly deuterated complex ions in the small quantity of unidentified peaks. This seems very unlikely, but is possible. 
\item The level of deuteration that has been achieved is less than equilibrium. 
\item Post-charging diffusion-loss of D through the surface of the specimen has occurred.
\item There is non-uniformity above the scale of the atom probe experiment
\item An unidentified APT-specific loss mechanism is in effect. 
Of note, these concentrations seem consistent with other works~\citep{Kesten2002}, where D/Metal ratios in a Fe/Pd/V multilayer reported D:V ratios of at most 0.18 (15at.\%), despite the pressure difference.  
\end{enumerate}

To examine the possibility of post-charging diffusion loss, a sample was stored for 6 days within the APT “buffer” chamber at high-vacuum conditions. Interestingly, the D content reported for this analysis, D:Pd = 0.025, was slightly higher than the D content reported for the analysis more immediately transferred, D:Pd = 0.022, despite both analyses being generated from same specimen.

Consequently, this work demonstrates that D is detectable after significant time, indicating that diffusion does not appear to be a major barrier to the study of D introduced into the Pd-Rh system. This is consistent with electrochemical results that exhibit a decreased effective diffusion rate for thin films, which is attributed to surface barrier effects~\citep{Li1996}, and bulk diffusion is reported elsewhere as not being the limiting factor in Pd-D diffusion~\citep{Schwarz2005}.

At room temperature, the sample should equilibrate purely within the $\beta$ phase. Upon cooling, or a reduction in D content consistent with observations as shown in Figure~\ref{fig:PdDProxigram}, the sample should transition into an $\alpha + \beta$ region. These two phases are both FCC but differ in lattice constant~\citep{Schwarz2005}. Thus, we anticipate that prior to charging the sample is $\alpha$, and then during charging it transitions to $\beta$, either partially or completely. Possibly there is some reversal during the experiment after the charging has been undertaken. It is possible that the enriched zone observed by APT is this $\beta$ phase.

\subsection{Vanadium}

Similarly to the Pd-Rh alloy, V also forms a hydride phase with H or D, and therefore this study also investigated the capability for APT to examine D using an ex-situ method. Previous work by the author has been reported on some preliminary in-situ low-pressure charging elsewhere, and low-pressure (hPa range) charging of V containing multilayers has been performed~\citep{Gemma2012,Kesten2002}.

\begin{figure}[htp]
 \centering
    \includegraphics[width=0.95\textwidth]{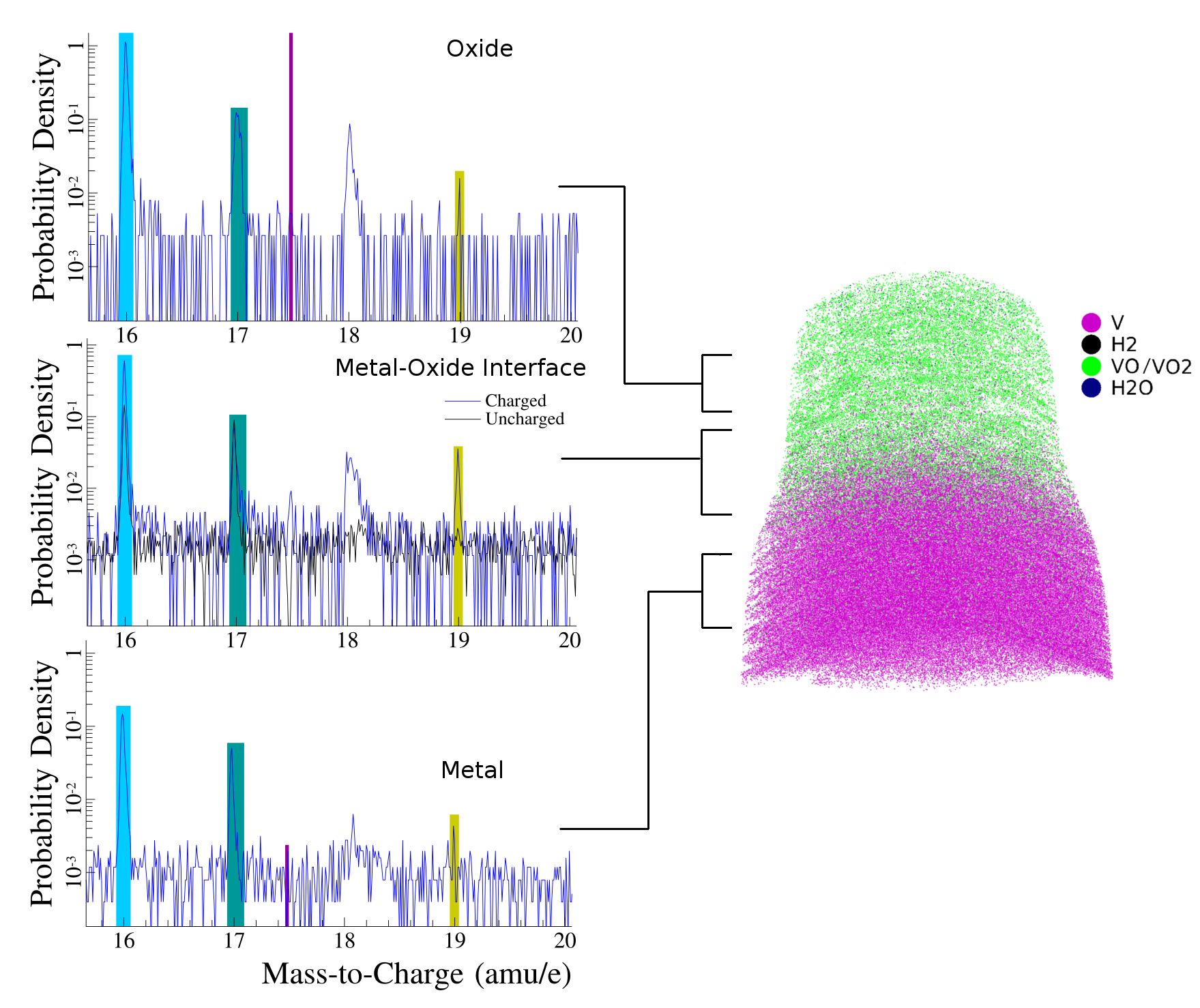}

 \caption{500~kPa (abs) Deuterium-charged Vanadium sample (voltage), showing section of mass spectrum at 3 different positions. Note the presence of a 19 peak at the metal-oxide interface, possibly $\mathrm{HDO}^{{}+{}}$.  Spectra normalised by total area. Uncharged spectrum from same tip shown as reference. V is not shown in the mass spectrum, as it occurs elsewhere (1 and 2+).}
 \label{fig:VandOxideInterface}
\end{figure}

The experiments in this study show the underlying metal as having no detectable deuterium uptake due to the formation of an oxide layer that occurs during venting and exposure to atmosphere. However, there is a change in the mass spectrum when traversing the oxide-metal boundary, which may be related to deuterium present at the metal-oxide interface. Figure~\ref{fig:VandOxideInterface} shows the mass spectra before, at and after analysis through the oxide region in the charged vanadium material (with uncharged for comparison). The peak at 19~Da appears enlarged as compared to either the charged bulk, oxide or the uncharged reference, suggesting the formation of $\mathrm{HDO}^{{}+{}}$ or alternately $\mathrm{H}_3\mathrm{O}^{{}+{}}$. If the former, this implies the presence of a small quantity of deuterium at the metal-oxide interface, and that some diffusion has occurred through the oxide barrier, which forms immediately on contact with air, thus this interface may therefore retain some deuterium. The presence of a water ion during an oxidation process of a readily oxidising hydride former is consistent  with the observations of Sundell~\citep{Sundell2016}. In the work of Sundell, a Zircaloy-2 material was charged with deuterium using an autoclave process. In  this work it was shown that there was segregation of $\mathrm{H}_2\mathrm{O}$ to grain boundaries that had been oxidised - thus it was demonstrated that this may be an oxidation pathway. By extension, it is possible that this small change in the 19~Da peak is a similar effect to that observed by Sundell at 18~Da, and reinforces the need for examining peaks beyond 2~Da. Differences are present between these experiments however, as in the work of Sundell a laser-mode analysis was used, where here a voltage study was performed, with differing target materials.

It is noted that the oxide kinetics are sufficiently fast that formation of a VO layer should have formed either within the atom probe load-lock, as it has been shown that exposure to air pressures greater than $10^{{}-5}$~hPa is sufficient to cause oxide formation on second timescales~\citep{Papathanassopoulos1982}.  

\subsection{Ex-situ charging in APT}

It is clear from these experiments that ex-situ charging can yield information on hydrides within atom probe, however there are several challenges that must be addressed. Primarily, maintaining surface cleanliness and limiting oxygen for rapidly oxidising samples is a key requirement. For more inert materials, ex-situ D charging can correctly identify D atoms within the metallic maltrix readily, and is a feasible route for hydride analysis. 

There remain several concerns for estimating the quantity of deuterium within the samples. Further work needs to be undertaken in order to fully understand the geometrical effect that arises from the use of needle samples, to which the reduced D content within these hydrides is attributed. Comparisons can be drawn with existing literature on hydrided samples, such as Nb measured by ERD~\citep{Romanenko2011,Maheshwari2011}.

\section{Conclusions}

In this work, we have demonstrated the capability of atom probe to perform ex-situ hydriding studies of bulk hydrides, in the form of PdRh-D. Within this material, a clear, spatially discriminated signal is observable on timescales that are, despite the small length scales inherent in APT, amenable to atom probe analysis. The stability of these hydrides was shown to be sufficient to allow for regular experimentation. However, there remains a discrepancy between the equilibrium solubilities that should be identifiable within this material, and the quantitative concentrations that have been observed within this work. For materials that rapidly form native oxides, it has been shown that the oxide layer itself may contain small quantities of deuterium, however an in-situ approach may be a more feasible alternative~\citep{Dumpala2015}, depending upon the propensity for the surface to oxidise.
Further work in this area should expand the pressure/temperature space that is experimentally available to atom probe, as well as to undertake multi-technique analyses, to fully understand the quantitative capacity of atom probe in deuterated studies.

\section{Acknowledgements}
The authors acknowledge the support of the EPSRC under the HEmS Programme Grant EP/L014742/1.

\bibliographystyle{MandM}
\bibliography{exsitucharging.bib}

\begin{table}
\begin{center}
\caption{Composition for uncharged state, as measured using voltage run. Data is average composition over three runs ($n=3$), error bounds are +- 2 standard error.}

\begin{tabular}{|l|p{3cm}|c|} \hline
Species & Mean Composition (At.\%), $n=3$ & $2 \times \mathrm{Std. Err}$\\ \hhline{|=|=|=|}
Pd & \centering 85.94 & 3.66\\\hline
Rh & \centering 8.81 & 0.11\\\hline
O & \centering 2.49 & 2.15\\\hline
H & \centering  2.31 & 1.27\\\hline
Unidentified & \centering 0.35 & 0.25\\\hline
N & \centering 0.09 & 0.04\\\hline
C & \centering 0.02 & 0.01\\\hline
\end{tabular}
\label{tbl:PdComposition}

\end{center}
\end{table}

\begin{table}
{%
\newcommand{\mc}[3]{\multicolumn{#1}{#2}{#3}}
\begin{center}
\caption{Estimated counts (background corrected) for uncharged and charged Pd-Rh alloy, (500~kPa, 1 hr charging) in the low-mass region of the dataset.}
\begin{tabular}{|p{1cm}|p{2cm}|l|lll}\cline{1-6}
\mc{2}{|c|}{\textbf{Uncharged State}} & \mc{4}{c|}{\textbf{Charged State}}\\\hline
Species & Backg. Corr. Counts & Species & \mc{1}{p{2cm}|}{Backg. Corr. Counts} & \mc{1}{p{1.7cm}|}{Estimated H  Counts} & \mc{1}{p{1.7cm}|}{Estimated D Counts}\\ \hhline{|=|=|=|=|=|=|}
H & \centering 2177.4 & H & \mc{1}{c|}{2428.5} & \mc{1}{c|}{-} & \mc{1}{c|}{×}\\\hline
$\mathrm{H}_2$& \centering 123.0 & D ${} + \mathrm{H}_2$ & \mc{1}{c|}{2477.6} & \mc{1}{c|}{137.2} & \mc{1}{c|}{2340.3}\\\hline
$\mathrm{H}_3$ & \centering 58.4 & DH ${} + \mathrm{H}_3$ & \mc{1}{c|}{166.2} & \mc{1}{c|}{65.1} & \mc{1}{c|}{101.0}\\\hline
$\mathrm{H}_4$ & \centering 0.0 & $\mathrm{D}_2$ & \mc{1}{c|}{40.9} & \mc{1}{c|}{0.0} & \mc{1}{c|}{40.9}\\\hline
\end{tabular}

\label{tbl:PdDCounts}
\end{center}
}%
\end{table}

\begin{table}
{%
\newcommand{\mc}[3]{\multicolumn{#1}{#2}{#3}}
\begin{center}
\caption{Relative fraction of D from each identified Pd-H/D species within the dataset ($1^{{}+{}}$ charge state), after resolving overlapped peaks. Surface ions were omitted from the analysis using a Pd${}>{}$50\% isosurface to exclude contamination.}
\begin{tabular}{l|c|ll}\cline{2-4}
{} & \mc{3}{c|}{\textbf{D${}_\mathbf{2}$ Charging Pressure (kPa, abs.)}}\\\cline{2-4}
{} & 5 & \mc{1}{c|}{500} & \mc{1}{c|}{200}\\\cline{2-4}
{} & \mc{3}{c|}{\textbf{Relative Species Fraction (At. Fr.)}}\\\hline
\mc{1}{|l|}{Pd} & 0.96 & \mc{1}{c|}{0.96} & \mc{1}{c|}{0.96}\\\hline
\mc{1}{|l|}{PdH} & 0.01 & \mc{1}{c|}{0} & \mc{1}{c|}{0}\\\hline
\mc{1}{|l|}{PdD} & 0.04 & \mc{1}{c|}{0.04} & \mc{1}{c|}{0.04}\\\hline
\mc{1}{|c|}{R${}^2$} & ${}>{}$0.999 & \mc{1}{c|}{${}>{}$0.999} & \mc{1}{c|}{${}>{}$0.999}\\\hline
\end{tabular}
\label{tbl:PdDFraction}
\end{center}
}%
\end{table}

\begin{table}
{%
\newcommand{\mc}[3]{\multicolumn{#1}{#2}{#3}}
\begin{center}
\caption{Relative composition of charged samples (At\%, background corrected, decomposed). Unidentified species have been treated as a single component for decomposition.}

\begin{tabular}{l|c|lll}\cline{2-5}
{} & \mc{4}{c|}{\textbf{Pressure (kPa, abs.)}}\\\cline{2-5}
{} & 500 & \mc{1}{c|}{500} & \mc{1}{c|}{500 (6 day delay)} & \mc{1}{c|}{200}\\\hline
\mc{1}{|l|}{\textbf{Species}} & \mc{4}{c|}{\textbf{Composition (At. \%)}}\\\hline
\mc{1}{|l|}{H} & 0.1 & \mc{1}{c|}{1.91} & \mc{1}{c|}{3.61} & \mc{1}{c|}{2.15}\\\hline
\mc{1}{|l|}{D} & 3.07 & \mc{1}{c|}{1.57} & \mc{1}{c|}{1.92} & \mc{1}{c|}{0.92}\\\hline
\mc{1}{|l|}{C} & 1.46 & \mc{1}{c|}{2.13} & \mc{1}{c|}{0.43} & \mc{1}{c|}{0.93}\\\hline
\mc{1}{|l|}{Unidentified} & 1.05 & \mc{1}{c|}{8.59} & \mc{1}{c|}{5.73} & \mc{1}{c|}{3.59}\\\hline
\mc{1}{|l|}{O} & 11.79 & \mc{1}{c|}{10.38} & \mc{1}{c|}{4.61} & \mc{1}{c|}{12.61}\\\hline
\mc{1}{|l|}{Rh} & 7.23 & \mc{1}{c|}{6.12} & \mc{1}{c|}{7.10} & \mc{1}{c|}{6.98}\\\hline
\mc{1}{|l|}{Pd} & 74.70 & \mc{1}{c|}{69.31} & \mc{1}{c|}{76.57} & \mc{1}{c|}{72.51}\\\hline
\end{tabular}
\label{tbl:ChargedPdComposition}

\end{center}
}%
\end{table}

\begin{table}
{%
\newcommand{\mc}[3]{\multicolumn{#1}{#2}{#3}}
\begin{center}
\caption{Composition of uncharged Vanadium samples, as measured by voltage APT. Oxide region has been previously removed by laser atom probe.}
\begin{tabular}{lll}\cline{2-3}
{} & \mc{2}{|c|}{Concentration, at\%}\\\hline
\mc{1}{|l|}{V} & \mc{1}{|c|}{87.16} & \mc{1}{|c|}{94.34}\\\hline
\mc{1}{|l|}{H} & \mc{1}{|c|}{10.37} & \mc{1}{|c|}{2.10}\\\hline
\mc{1}{|l|}{O} & \mc{1}{|c|}{2.13} & \mc{1}{|c|}{3.46}\\\hline
\mc{1}{|l|}{Si} & \mc{1}{|c|}{0.16} & \mc{1}{|c|}{-}\\\hline
\mc{1}{|l|}{N} & \mc{1}{|c|}{0.074} & \mc{1}{|c|}{0.060}\\\hline
\mc{1}{|l|}{C} & \mc{1}{|c|}{0.067} & \mc{1}{|c|}{-}\\\hline
\mc{1}{|l|}{Unidentified} & \mc{1}{|c|}{0.044} & \mc{1}{|c|}{0.031}\\\hline
\end{tabular}
\label{tbl:ChargedVComposition}
\end{center}
}%

\end{table}

\clearpage

\end{document}